\documentclass[usenatbib]{mnras}
\usepackage{amssymb}

\usepackage[T1]{fontenc}

\DeclareRobustCommand{\VAN}[3]{#2}
\let\VANthebibliography\thebibliography
\def\thebibliography{\DeclareRobustCommand{\VAN}[3]{##3}\VANthebibliography}

\usepackage{bm}
\usepackage{graphicx}	

\title[Nuclei finite size effect for shear modulus]
{Neutron star inner crust: reduction of shear modulus by nuclei finite size effect}

\author[N.\ A.\ Zemlyakov, A.\ I. Chugunov]{
	Nikita A.\ Zemlyakov$^{1,2}$\thanks{E-mail: zemnic5@gmail.com}
	Andrey I.\ Chugunov,$^{1}$\thanks{E-mail: andr.astro@mail.ioffe.ru}
	\\
	$^{1}$Ioffe Institute, Politekhnicheskaya 26, 194021 St. Petersburg, Russia \\
	$^{2}$Peter the Great St.\ Petersburg Polytechnic University, Politekhnicheskaya 29 , St.\ Petersburg 195251, Russia;
}

\date{Accepted XXX. Received YYY; in original form ZZZ}

\pubyear{2022}

\begin{document}
	\label{firstpage}
	\pagerange{\pageref{firstpage}--\pageref{lastpage}}
	\maketitle

\begin{abstract}
The elasticity of neutron star crust is important for adequate interpretation of observations. To describe elastic properties one should rely on theoretical models. The most widely used is Coulomb crystal model (system of point-like charges on neutralizing uniform background), in some works it is corrected for electron screening. These models neglect finite size of nuclei. This approximation is well justified except for the innermost crustal layers, where nuclei size becomes comparable with the inter-nuclear spacing. Still, even in those dense layers it seems reasonable to apply the Coulomb crystal result, if one assumes that nuclei are spherically symmetric: Coulomb interaction between them should be the same as interaction between point-like charges. This argument is indeed correct, however, as we point here, shear of crustal lattice generates (microscopic) quadrupole electrostatic potential in a vicinity of lattice cites, which induces deformation on the nuclei. We analyze this problem analytically within compressible liquid drop model, using ionic spheroid model (which is generalization of well known ion sphere model). In particular, for ground state crust composition the effective shear modulus is reduced for a factor of $1-u^{5/3}/(2+3\,u-4\,u^{1/3})$, where $u$ is the filling factor (ratio of the nuclei volume to the volume of the cell). This result is universal and does not depend on the applied nucleon interaction model. For the innermost layers of inner crust $u\sim 0.2$ leading to reduction of the shear modulus by $\sim 25\%$, which can be important for correct interpretation of quasi-periodic oscillations in the tails of magnetar flares.
\end{abstract}


\section{Introduction}

The core of neutron stars is generally believed to be composed of microscopically uniform nuclear (or quark) matter (e.g., \citealt*{hpy07}). However, it is not the case for the crust, where nuclear matter is clumped into clusters (nuclei, ions), located on the background of almost uniform degenerate electron  gas and, except for the outer crust, additional background of unbound neutrons (e.g., \citealt{ch08}). For typical temperatures of neutron stars ($T<10^9$~K) the Coulomb interaction between  these nuclei is much larger than the thermal energy and nuclei become ordered into a lattice and crust solidifies.
As for terrestrial conditions, solidification allows for elastisity: solid crust  supports shear stresses, and,  
according to the state-of-art models, this effect indeed affects observations. Crustal elasticity is supposed to be responsible for quasi-periodic oscillations after giant flares of magnetars (e.g., \citealt{hc80_torsOsc,st83_tors_osc,McDermott_ea88,Strohmayer_etal91,Gabler_ea11,Gabler_ea12,Gabler_ea13,Gabler_ea18,Sotani_ea18,KY20}), 
it can support static  asymmetry in the mass distribution in the crust (mountains), which can lead to emission of gravitational waves (see e.g., \citealt{Ushomirsky_ea00,hja06_mountains,Horowitz10_lowmass,McDaniel_Owen13,gaj20,KM22_Mountain,mh22}
for the models and \citealt{LIGO_VIRGO20_MSPelipt,LIGO21_Pulsars,LIGO22_GW_isotated} for recent observational constraints).

Assuming that the matter in the neutron star crust is macroscopically isotropic (e.g., it is polycrystalline), the coupling of the shear stress with deformation can be described by the shear modulus,
which should be calculated theoretically (the matter of neutron star crust cannot be obtained and directly studied in the laboratory).
This problem were analyzed  in a long list of papers (e.g., \citealt{Strohmayer_etal91,hh08,hk09,Baiko11,Baiko12,kp15,Baiko15,Kozhberov19_elast,C21_elastCoins, Kozhberov22,Chugunov22_elast_screen}). 
Typically, these works replace nuclei by point-like massive particles with Coulomb or screened Coulomb (Yukawa) interaction, arrange these particles into lattice (or polycrystalline system)
and consider energy change (or stresses) which arises in this system as a result of deformation.
The results for static one-component Coulomb crystals with body- and face-centred lattices were obtained almost a century ago by \cite{Fuchs36} and more recent works improve them, analyzing also multicomponent lattices and screening (e.g.\ \citealt{Kozhberov19_elast}, see also  \citealt{C21_elastCoins,Chugunov22_elast_screen} for recent approximate, but universal formulae), as well as quantum and thermal motion of the particles (\citealt{Strohmayer_etal91,hh08,Baiko12}).

To replace nuclei with point-like massive particles seems very natural approximation, at least if nuclei sizes are much smaller than internuclear distances.
It is indeed true for the major part of the crust, except for the innermost crustal layers, where nuclei (proton) radius $R$
can exceed half of the nuclei cell radius $a=\left(4\pi n_\mathrm N/3\right)^{-1/3}$ (note, the distance to the nearest neighbour for body-centered cubic lattice is $(n_\mathrm N/2)^{-1/3}\approx 2 a$). Here $n_\mathrm N$ is number density of nuclei.
However, if one one neglects electron screening and assumes that nuclei charge (proton) density is spherically symmetric, the electrostatic theory suggests that Coulomb  interaction energy of two nuclei should be  the same as for the point-like charges (up to effects associated with overlap of proton density profiles).
Thus, the shear modulus, determined by energy change associated with deformation, should be unaffected by finite nuclei size, being the same as for Coulomb crystal, if nuclei remain spherically symmetric during deformation. 
Here we demonstrate that the latter assumption generally is not true, if one considers deformation of neutron star crust.

Namely, even if nuclei are spherically symmetric for undeformed crust, deformation would induce the quadrupole asymmetry. Taken into account, this effect reduces the energy in deformed state (in the opposite case, it is not energetically favourable and would not occur) and, thus, the shear modulus.

In this paper we limit ourselves to consideration of the toy problem, which is based on compressible liquid drop model (CLDM) with approximation of primitive lattice cell in undeformed state by a sphere (e.g.\ supplementary material of \cite{GC20_DiffEq} for description of the applied CLDM).
Namely, we model infinitesimal  deformation of the crust by respective deformation of the spherical cell, which becomes a spheroid (ion spheroid model).
We apply CLDM to calculate energy of this spheroidal cell, allowing for deformation of nucleus in the center of the cell.
The deformation parameter of the cell $\varepsilon$ serves as a driving parameter for nuclei deformation $\varepsilon_\mathrm{p}$.
In this way we demonstrate that:
(1) If nucleus remains spherical ($\varepsilon_\mathrm{p}=0$), our model reproduces shear modulus, as obtained by \cite{C21_elastCoins} using ion sphere approximation. It serves as justification of our model;
(2) Minimization of the cell energy leads to $\varepsilon_\mathrm{p}\propto \varepsilon$ and reduces shear modulus by a factor $1-u^{5/3}/(2+3\,u-4\,u^{1/3})$ in the case of the ground-state crust.
The maximal reduction by $\sim 25\%$ takes place at the inner layers of the neutron star crust, where the filling factor is maximal $u\sim 0.2$. As shown by \cite{KY20}, these layers are the most important for torsional modes oscillation spectrum, and therefore it is crucial to know the shear modulus accurately to interpret observed quasi-periodic oscillation of magnetars correctly.

\section{Crust elasticity within ion spheroid model}
Generally, elastic properties can be calculated by considering the energy change associated with deformation of solid. If one neglects nuclei motion and consider static lattice, it is enough to calculate the energy change of primitive cell, imposing appropriate boundary conditions (for Coulomb crustal is it periodic boundary conditions for electrostatic potential).
Here, as discussed in the introduction, we apply a toy model. Namely, we follow  a well-known spherical Wigner-Seitz cell approximation and replace the primitive cell in undeformed state (truncated octahedron in the case of body-centered cubic crystal) by a sphere.
After that, we apply deformation for this spherical cell and calculate energy change associated with deformation.
To justify this approach, we demonstrate that it reproduces the results of accurate calculations of effective shear modulus for Coulomb crystal well enough.
We also discuss the approaches, which can be applied to make accurate calculations in the last section (Summary and discussion).

To calculate the energy of deformed cell we generalize  CLDM by \cite{GC20_DiffEq}.
The first step was done in \cite{ZC22_stability}, where  we consider deformed (spheroidal) nucleus in spherical cell and conclude that spherical nuclei should be stable with respect to infinitesimal deformation, if their number density $n_\mathrm N$ is not too small for a given baryon number density (in particular, spherical nuclei in ground state and accreted crust was shown to be stable with respect to considered deformation  at arbitrary filling factor).
Here we make the next step and assume infinitesimal deformation of the initially spherical cell.

As long as our model problem is spherically symmetric in undeformed state, its elastic properties can be described with just two constants: bulk modulus $K$ and shear modulus $\mu$. Bulk modulus $K$ is determined by the equation of state, and therefore is generally well-known. Here we consider only $\mu$ and determine it, applying volume-conserving deformation,%
\footnote{Such method helps to avoid complications of finite pressure elasticity theory, see \cite{Chugunov22_elast_screen} for brief discussion and references for details.}
which deforms cell to the spheroid with axis $a(1+\varepsilon)$ and $a/\sqrt{1+\varepsilon}$. 
According to definition of $\mu$, this deformation should change the cell energy by the amount 
\begin{equation}
\delta E=\frac{3}{2} \mu \varepsilon^2 V_\mathrm c,
\label{dE_mu}
\end{equation}
 where cell volume $V_\mathrm c =1/n_\mathrm N=4\pi a^3/3$.
Below we calculate the energy change $\delta E$ and apply (\ref{dE_mu}) to quantify  $\mu$.

Within our CLDM the nucleus is assumed to be a spheroid with axes $R(1+\varepsilon_\mathrm{p})$ and $R/\sqrt{1+\varepsilon_\mathrm{p}}$, which is coaxial with the cell (note, parameter $\varepsilon_\mathrm{p}$ does not affect the nucleus volume).
%
As a result, the parameter space of CLDM by \cite{GC20_DiffEq} [neutron and proton number density  inside nucleus ($n_\mathrm{ni}$ and $n_\mathrm{pi}$ respectively), neutron number density outside nucleus ($n_\mathrm{no}$; we assume that protons are absent outside nucleus), surface density of adsorbed neutrons $\nu_\mathrm{s}$,  nucleus (proton) radius $R$,  and cell volume $V_\mathrm{c}=4\pi a^3/3\equiv n_\mathrm{N}^{-1}$] is supplemented with two additional parameters: $\varepsilon$ and $\varepsilon_\mathrm{p}$.
Within this paper these two are treated as infinitesimal.
To make the equations compact, it is useful to use total baryon number density  $n_\mathrm{b}$  instead of $\nu_\mathrm{s}$, filling factor $u=R^3/a^3$ instead of $R$, and $n_\mathrm{N}$ instead of $V_\mathrm{c}$.
Denoting resulting parameter set of spherically symmetric CLDM as $X=\{n_\mathrm{b}, n_\mathrm{N}, n_\mathrm{ni}, n_\mathrm{pi},n_\mathrm{no},u\}$, we write down the cell energy (see Appendix for derivation details):
\begin{equation}
E=E^0(X)+f_1(X)\varepsilon^2+f_2(X)\varepsilon_\mathrm{p}^2+f_3(X)\varepsilon\varepsilon_\mathrm{p},
\label{E}
\end{equation}
where $E^0(X)$ is the energy of spherical cell with spherical nucleus (see \citealt{GC20_DiffEq}),
\begin{eqnarray}	
f_1&=&\frac{9}{50}\frac{Z^2 e^2}{a},\label{f1}\\
f_2&=&\frac{8}{5}\pi R^2\sigma+\frac{3}{50}\frac{Z^2 e^2}{R}(5u-2),
\label{f2}\\
f_3&=&-\frac{9}{25}\frac{Z^2 e^2}{R}u.\label{f3}
\end{eqnarray} 
The first term in the right-hand side of (\ref{f2}) corresponds to the change of the surface energy ($\sigma$ is surface tension), associated with nucleus deformation, while the second is related to the change of the Couloumb energy. Note, that the latter has a correction with respect to a well-known result for isolated nucleus (e.g., \citealt{BW39_fission})
associated with the presence of the neutralizing electron background in the cell (see \citealt{ZC22_stability} for more detailed discussion).

For completeness we do not assume ground state crust composition and allow for non-equlibrium number density of nuclei $n_\mathrm N$.
As discussed by \cite{GC20_DiffEq}, it can be the case, e.g., for accreted crust.

First, let us consider a simplified problem and assume that the nucleus is spherical, $\varepsilon_\mathrm{p}=0$. 
The parameter set $X$ should be determined by  minimization of energy at a given baryon number density $n_\mathrm{b}$, nuclei number density $n_\mathrm{N}$, and deformation $\varepsilon$.
As long as $\varepsilon$ is infinitesimal we can present $X$ as $X=X^0+\delta X$, where $X^0$ corresponds to undeformed crust: they are energetically optimal parameters set for the same $n_\mathrm{b}$ and $n_\mathrm N$, but $\varepsilon=0$. It is easy to show that $\delta X\propto \varepsilon^2$.
Since $X^0$ corresponds to a minimum of $E^0$, up to the second order in $\varepsilon$ the energy change associated with deformation is $\delta^\mathrm{sp} E=E(X,\varepsilon_\mathrm{p}=0,\varepsilon)-E(X^0,\varepsilon_\mathrm{p}=0,\varepsilon=0)=f_1(X^0)\varepsilon^2$ (here, the upper index `sp' indicates that it is obtained assuming nuclei to have spherical shape).
Using Eq.\ (\ref{dE_mu}), we obtain 
\begin{equation}
\mu^\mathrm{sp}=\frac{2}{3V_c} f_1=\frac{3}{25}\frac{Z^2 e^2}{a}n_\mathrm N.
\label{mu_sp}
\end{equation}
This result coincides with the estimate obtained by \cite{C21_elastCoins} using ion sphere model, and, as pointed in that work,  it differs only by $\sim 0.5\%$ from the accurate calculations of Voigt averaged shear modulus.
It is also worth noting, that, following the discussion in the introduction, this result does not depend on the  filling factor $u$ and neither on nuclei size.

Let us now get rid of  $\varepsilon_\mathrm{p}=0$ assumption and treat it as a free parameter.
In this case, not only parameters $X$, but also $\varepsilon_\mathrm{p}$ should be determined by minimization of energy at a given baryon number density $n_\mathrm{b}$, nuclei number density $n_\mathrm N$, and deformation $\varepsilon$. For $\varepsilon_\mathrm{p}$ it leads
\begin{equation}
\varepsilon_\mathrm{p}=-\frac{f_3(X)}{2f_2(X)}\varepsilon
=\frac{u}{2-4u^{1/3}+3u+\tilde \eta_\mathrm N}\varepsilon,
\label{epsilon_p}
\end{equation}
where $\tilde \eta_\mathrm N =20\eta_\mathrm N R/(3 Z^2 e^2)$ and  $\eta_\mathrm N$ is energy, required to create an additional cluster from existing baryons ($\eta_\mathrm N=0$ for ground state matter, i.e.\ equilibrium value of $n_\mathrm N$).
The term $\tilde \eta_\mathrm N$ arises in the denominator of (\ref{epsilon_p}) as a result of exclusion of surface tension using relation for undeformed crust
%
\begin{equation}
3\eta_\mathrm N=4\pi \sigma R^2-\frac{6}{5}\frac{Z^2 e^2}{R}\left(1-\frac{3}{2}u^{1/3}+\frac{1}{2}u\right).
\label{vir}
\end{equation}
It is generalization of the virial theorem and allows to calculate  $\eta_\mathrm N$ for crust with non-equilibrium $n_\mathrm N$ (all other parameters of CLDM are adjusted to minimize energy, which leads to set of equations that can be interpreted as beta-equilibrium, chemical and mechanical equilibrium conditions for the cell, see supplementary material in \citealt{GC20_DiffEq} for details).%
\footnote{\cite{GC20_DiffEq} denote $\eta_\mathrm N$ as $\mu_\mathrm N$, but we do not follow this notation to avoid confusion with shear modulus, also denoted as $\mu$.}

As for the $\varepsilon_\mathrm{p}=0$ case discussed above, $X$ can be presented as $X^0+\delta X$, where $\delta X\propto \varepsilon^2$. Substituting (\ref{epsilon_p}) into (\ref{E}) and  keeping only the second-order terms in $\varepsilon$, it is straightforward to write down an explicit expression for $\mu$ using (\ref{dE_mu}):
\begin{equation}
\mu=\frac{3}{25}\frac{Z^2 e^2}{a}n_\mathrm N\left(1-\frac{u^{5/3}}{2-4u^{1/3}+3u+\tilde \eta_\mathrm N}\right).
\label{mu}
\end{equation}
This equation differs from the result for the point-like nuclei (Eq.\ \ref{mu_sp}) only by the  factor in the brackets, which is referred below as `correction factor' and denoted $C\equiv 1-u^{5/3}/(2-4u^{1/3}+3u+\tilde \eta_\mathrm N) $.


\begin{figure}                                           
	\begin{center}                                              
		\leavevmode                                                 
		\includegraphics[width=\columnwidth]
		{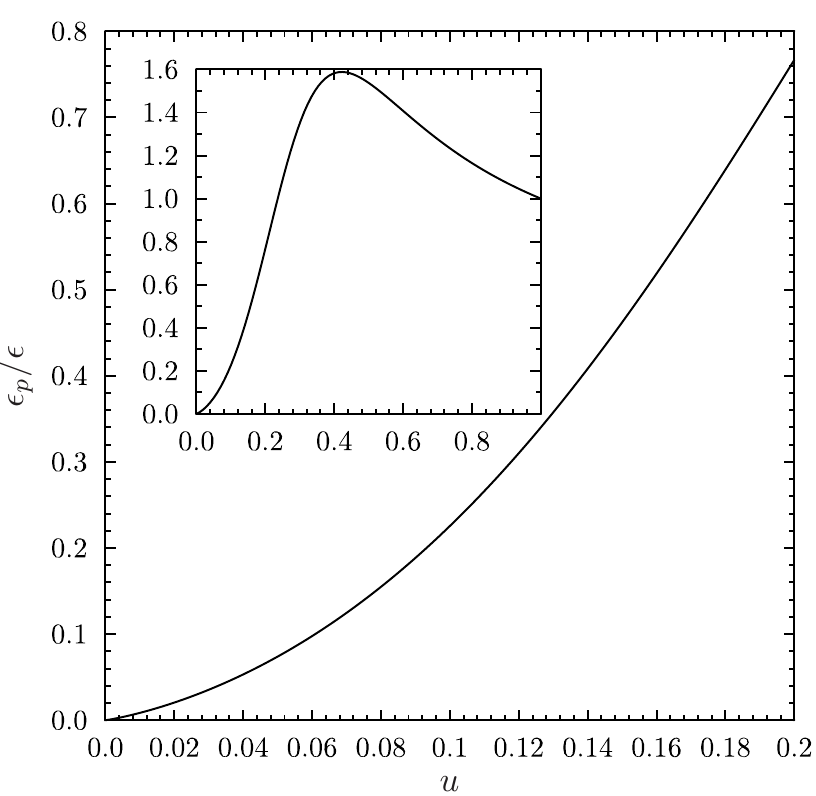} 
	\end{center}                                                
	\caption{$\varepsilon_\mathrm{p}/\varepsilon$ ratio as function of filling factor $u$ for ground-state crust.
	At the main plot $u$ is  limited to physically motivated region $u\le0.2$. The inset represents the same function for $u\le 1$.}                                             
	\label{Fig_eps_ratio}
\end{figure}
%

\begin{figure}                                           
	\begin{center}                                              
		\leavevmode                                                 
		\includegraphics[width=\columnwidth]
		{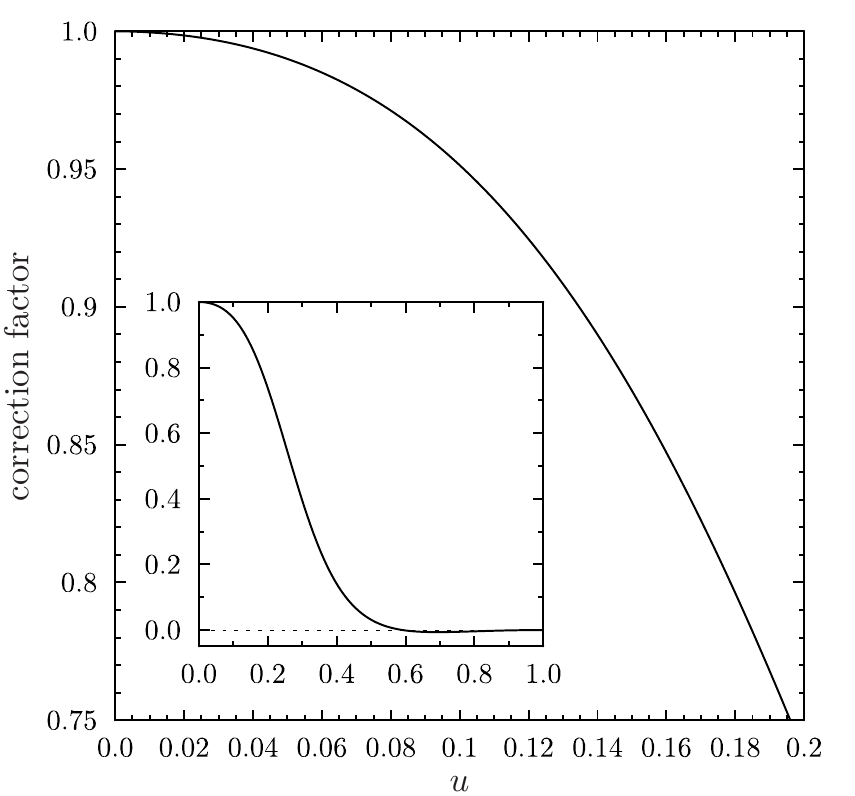} 
	\end{center}                                                
	\caption{Correction factor for the shear modulus as function of filling factor $u$ for ground state crust for  physically motivated region $u<0.2$.
		The inset represents correction factor for $u\le 1$
	}                                             
	\label{Fig_mu_cor}
\end{figure}
%

The ratio $\varepsilon_\mathrm{p}/\varepsilon$ is shown in Fig.\ \ref{Fig_eps_ratio} for ground-state crust ($\eta_N=0$) and physically motivated region  $u\le0.2$.
It is worth to stress that
this figure does not depend on the microphysical model, applied to quantify parameters of CLDM (bulk energy density and surface tension).
One can see that the  $\varepsilon_\mathrm{p}/\varepsilon$ ratio increases monotonically from zero at $u=0$.
It confirms that for small filling factors nucleus deformation is negligible. For $u=0.2$ the ratio $\varepsilon_\mathrm{p}/\varepsilon$ reaches $\approx 0.75$.  
For the sake of completeness, the inset demonstrates $\varepsilon_\mathrm{p}/\varepsilon$ ratio for $u\le 1$.
In this region the ratio becomes non-monotonic  and has a maximum at filling factor $u\sim 0.4$, where deformation of nucleus $\varepsilon_\mathrm{p}$ is larger than the crust deformation $\varepsilon$. However, it is unlikely that there are any nuclei with so large filling factors in realistic models of the crust.

The dependence of the shear modulus correction factor $C$
on $u$ is shown in Fig.\ \ref{Fig_mu_cor}, taken ground-state crust as an example ($\tilde \eta_\mathrm N=0$).
As for Fig.\ \ref{Fig_eps_ratio}, this plot does not depend on the numerical parameters of CLDM.
As expected, $C\approx 1$ (no correction), if nuclei size is negligible ($u\ll1$), but for $u\sim0.1$ correction factor for nuclei deformation becomes noticeable ($C\approx 0.95$) and for $u=0.2$,  which is typical for the innermost crystal layers, where spherical nuclei are present (e.g., \citealt{ZC22_stability} and references therein),  shear modulus is reduced by $25\%$ ($C\approx 0.75$). 
The inset demonstrates the behaviour of the correction factor in the whole mathematically possible region, up to $u=1$. One can see that the correction factor becomes negative for $u\gtrsim 0.6$, however filling factors that large are not expected for the crust and our model is possibly oversimplified to analyze this region.

\begin{figure}                                           
	\begin{center}                                              
		\leavevmode                                                 
		\includegraphics[width=\columnwidth]
		{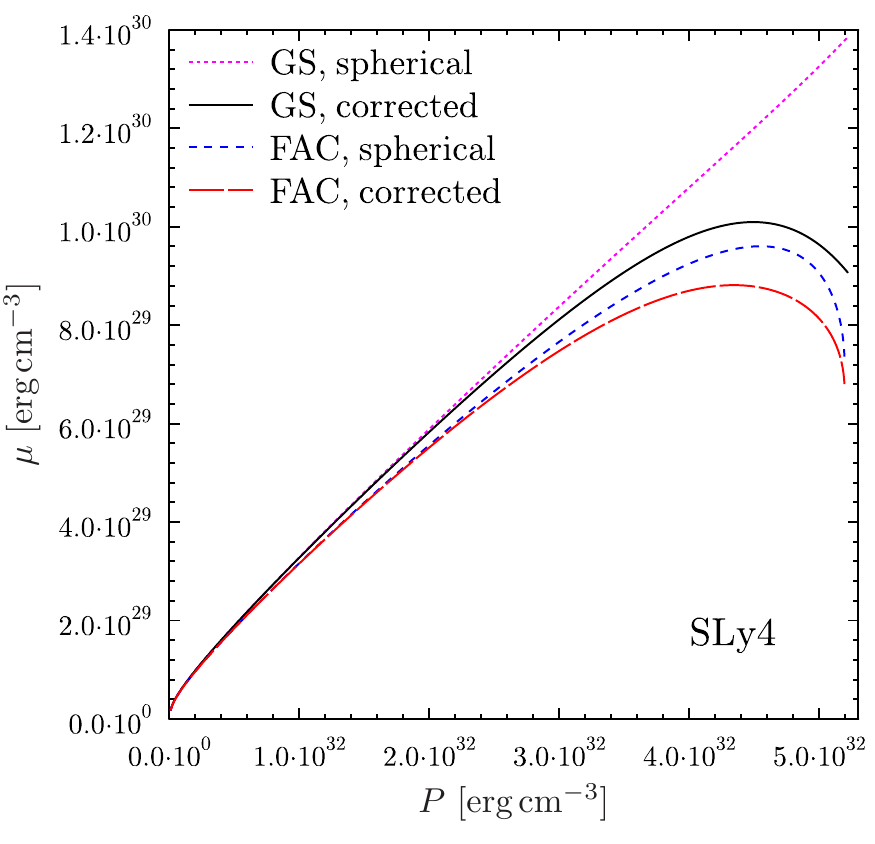} 
	\end{center}                                                
	\caption{Shear modulus of the densest region of inner crust as function of pressure $P$.
	Dotted line and short dashes represent shear modulus neglecting correction for nuclei deformation ($\mu^\mathrm{sp}$ as given by Eq.\ \ref{mu_sp}) for ground state (GS) crust and fully accreted crust (FAC), respectively.
	Solid and long dashes represent shear modulus, corrected for nuclei deformation according to Eq.\ (\ref{mu}) for GS and FAC, respectively.
	Numerical results corresponds to SLy4 energy-density functional, shell effects were neglected.
	The lines end at the crust-core interface	
	}                                             
	\label{Fig_mu}
\end{figure}
%

To illustrate the effect of non-equilibrium nuclei number density $n_\mathrm{N}$ we take fully accreted crust model by  \cite{GC20_DiffEq} as an example.
This model takes into account neutron hydrostatic/diffusion (nHD) equlibrium condition and
based on SLy4 nucleon-nucleon potential (\citealt{Chabanat_ea98_SLY4,Chabanat_ea98_SLY4_p2_nuclei}).
The model neglects shell effects, being thus uniquely specified for chousen set of CLDM microphysical parameters (\citealt{GC20_DiffEq,GC21_HeatReleaze}).
Resulting dependencies of the shear modulus on pressure for ground state and fully accreted crust models are shown in Fig.\ \ref{Fig_mu}. 
High-pressure end of lines corresponds to the crust-core interface (see \citealt{GC20_DiffEq} for details).

If nuclei deformation is neglected, shear modulus $\mu=\mu^\mathrm{sp}$ is predicted to be almost linear function of pressure for ground state (GS) crust composition (dotted line in Fig.\ \ref{Fig_mu}).
If nuclei deformation is taken into account, the shear modulus decreases substantially (solid line at Fig.\ \ref{Fig_mu}).

For fully accreted crust
model by \cite{GC20_DiffEq}, the nuclei charge $Z$ is just a bit lower than  for ground-state crust up to the densest regions of inner crust, where the difference starts to increase (see Fig.\ 2 there). As a result, shear modulus of accreted crust is almost the same as for the ground-state crust up to $P\approx 10^{32}$~erg\,cm$^{-3}$, but becomes noticeably lower for larger pressure, even if nuclei deformation is neglected (short dashed line at Fig.\ \ref{Fig_mu}). Account for nuclei deformation reduces the shear modulus, but this effect is weaker than for ground state crust. It happens because for fully accreted crust  $\tilde \eta_\mathrm N>0$ and reduces a correction factor according to Eq.\ (\ref{mu}).

It should be noted, that shell effects, which are not now taken into account here, can modify composition of fully accreted crust substantially, leading to even lower nuclei charges (see \citealt{GC21_HeatReleaze}).
As a result, the shear modulus of accreted crust can be substantially reduced.

\section{Summary and discussion}
We analyze nuclei finite size effects for shear modulus of neutron star crust and conclude that they become non-negligible for the innermost layers, where the filling factor $u\sim 0.1$.
Our results are based on CLDM and spheroidal primitive cell approximation for the deformed crustal lattice. This model allows one to analyze the problem analytically and obtain a general expression for the correction factor for effective shear modulus $C$ (Eq.\ \ref{mu_sp}). This correction reduces the shear modulus ($C<1$).
In the case of ground-state composition of the crust, which is generally believed to be a reasonable approximation for isolated neutron stars (e.g., \citealt{hpy07}, see, however, \citealt{cfg20,pc21}),  it has especially simple form $C=1-u^{5/3}/(2+3\,u-4\,u^{1/3})$.

The reduction of the shear modulus has simple physical nature: deformed lattice induces asymmetric quadrupolar electrostatic field in a vicinity of lattice sites, where nuclei are located. Similarly to the tidal field in the case of neutron star mergers or Earth's tides, the nuclei deformation becomes energetically favourable in the presence of this field. Accounting for this effect 
decreases the energy in the deformed state and thus estimate for shear modulus is reduced.

Clearly, this work presents a simplified model. An accurate calculations should include precise description of the induced quadrupole electric field, which generally depends on the lattice type as well as on the type of deformation. Detailed (quantum) description of the nuclear response for this field is also required.
However, accurate treatment for infinitesimal deformations should lead to energy dependence in form of Eq.\ (\ref{E}). We believe that our model gives a reasonable (at least by order of magnitude) estimate for parameters of this equation: $f_1$, $f_2$, and $f_3$, and thus for the final result.
Namely, for $f_1$, which is determined by deformation of lattice with spherical (or point-like) nuclei we have a good approximation of accurately calculated Voigt averaged effective shear modulus.
The parameter $f_2$, which describes energy change associated with nuclei deformation, is estimated within CLDM model, which should be reasonable for nuclei with large number of nucleons.
Finally, for parameter $f_3$, which is responsible for coupling of lattice and nucleus deformation, we obtain an estimate with natural dependence on the parameters (filling factor, nuclei charge, and radius), which should give a correct order of magnitude value.
We plan to get rid of a spheroidal cell approximation and analyze nuclei finite size effects for realistic lattice types in subsequent publications.

Please, be aware that our calculations neglect electron screening and nuclei motion.
The latter effects were analyzed in literature only in approximation of  point-like particles (e.g., \citealt{Baiko12}). 
However, for practical applications one needs an approach, which combines all necessary  effects.
For this aim we would like to suggest to multiply the results of \cite{Baiko12} (or, e.g., \citealt{oi90} supplemented with the electron screening correction from \citealt{Chugunov22_elast_screen}) by the correction factor  (\ref{mu_sp}).
It would not affect shear modulus in the outer crustal region, where point-like approximation is applicable, while in the innermost crustal regions the shear modulus will be corrected for nuclei finite size effects.
Note, that the melting temperature for the innermost layers of neutron star crust is expected to be much larger than the typical neutron star temperatures and nuclei motion effects can be neglected. 
Finally, the suggested approach would protect against artificial instabilities ($\mu<0$), which potentially can arise at extreme parameters, if electron screening and nuclei motion corrections are applied to reduce shear modulus, given by Eq.\ (\ref{mu}).
	
\section*{Acknowledgements}
The work of AIC was partially supported
by The Ministry of Science and Higher
Education of the Russian Federation (Agreement with Joint Institute for High Temperatures RAS No 075-15-2020-785 dated September 23, 2020).	
\section*{DATA AVAILABILITY}
The data underlying this paper are available in the paper.

\bibliographystyle{mnrass}

\bsp	

\section*{Appendix}
\label{app}

Derivations in the paper are based on generalization of CLDM by \cite{GC20_DiffEq}, which incorporates surface tension $\sigma$ thermodynamically consistently,  neglecting curvature effects.
Here, instead of spherical cell with spherical nucleus at the cell center, considered by \cite{GC20_DiffEq},
we study spheroidal cell with spheroidal nucleus (both spheroids assumed to be coaxial).%
\footnote{We checked that more general treatment, allowed for non-coaxial orientation, is not required because coaxial orientation of nucleus minimizes energy at least for considered CLDM.}
The semi-axes of the cell are  $a(1+\varepsilon)$ and $a/\sqrt{1+\varepsilon}$, while for the nucleus they are  $R(1+\varepsilon_\mathrm{p})$ and $R/\sqrt{1+\varepsilon_\mathrm{p}}$. 
The deformation parameters $\varepsilon$ and $\varepsilon_\mathrm{p}$ do not affect cell and nucleus volume, and thus the bulk terms of CLDM (the first line in Eq.\ \ref{Ecldm}) are also unaffected.
As a result,  the energy density can be written as:
\begin{eqnarray}
\epsilon&=&u\,\epsilon^\mathrm{bulk}(n_\mathrm{ni},n_\mathrm{pi})
+(1-u)\,\epsilon^\mathrm{bulk}(n_\mathrm{no},0)+\epsilon_\mathrm{e}(n_\mathrm{e})
\label{Ecldm}
\\
&+&\frac{E_\mathrm{s}(\nu_\mathrm s, R,\varepsilon_\mathrm{p}, \varepsilon)}{V_\mathrm{c}}
+\frac{E_\mathrm{C}(n_\mathrm{pi}, R,u,\varepsilon_\mathrm{p}, \varepsilon)}{V_\mathrm{c}}
.\nonumber
\end{eqnarray}
Here $\epsilon^\mathrm{bulk}(n_\mathrm{n},n_\mathrm{p})$ is the energy density of bulk nuclear matter at respective neutron and proton number density ($n_\mathrm{n}$ and $n_\mathrm{p}$),
$n_\mathrm{ni}$, $n_\mathrm{pi}$ are the neutron and proton density inside nucleus, 
$n_\mathrm{no}$ is the neutron density outside nucleus (we assume that there is no proton drip), and $\epsilon_\mathrm{e}(n_\mathrm{e})$ is the energy density of degenerate electrons at electron number density $n_\mathrm{e}=u\,n_\mathrm{pi}$.
The cell volume $V_\mathrm{c}=4\pi a^3/3=4\pi R^3/(3u)$.
$E_\mathrm{s}(\nu_\mathrm s, R,\varepsilon_\mathrm{p}, \varepsilon)$ and $E_\mathrm{C}(n_\mathrm{pi}, R,u,\varepsilon_\mathrm{p}, \varepsilon)$ are the surface and the Coulomb energies of the cell, which are calculated below, assuming that the deformation parameters $\varepsilon$ and $\varepsilon_\mathrm{p}$ are infinitesimal, being of the same order ($\varepsilon~\sim~ \varepsilon_\mathrm{p}$.).
Only the lowest order non-trivial terms are kept below.

Let's start from the the surface energy $E_\mathrm{s}$.
As long as our CLDM neglects curvature corrections, but incorporates neutron adsorption, the surface energy can be written as (e.g., \citealt{ll80,lpr85}):
\begin{equation}
E_\mathrm{s}=\mathcal A \left(\mu_\mathrm{ns}\nu_ s+\sigma\right).
\end{equation} 
Here 
the nuclei surface area $\mathcal A$ is given by the surface area of the spheroid:
\begin{equation}
\mathcal A =4\pi R^2+\frac{8}{5}\pi R^2 \varepsilon_\mathrm{p}^2,
\end{equation} 
while $\mu_\mathrm{ns}$ and $\nu_\mathrm s$ are chemical potential and surface density of the adsorbed neutrons.
Thermodynamic consistency requires (e.g., \citealt{ll80,lpr85}):
\begin{equation}
\frac{d \sigma}{d \nu_\mathrm{s}}=-\nu_\mathrm{s} \frac{d \mu_\mathrm{ns}}{d \nu_\mathrm{s}},
\label{ThermCons_sigma}
\end{equation}
thus $\sigma$ and $\mu_\mathrm{ns}$ can be treated as functions of $\nu_\mathrm{s}$.

The parameter set, applied in this work, $X\equiv\{n_\mathrm{b},n_\mathrm{N},n_\mathrm{ni}, n_\mathrm{pi},n_\mathrm{no},u\}$, $\varepsilon$, $\varepsilon_\mathrm{p}$. It doesn't use $\nu_\mathrm{s}$ explicitly, and it should be calculated  using relation
%
\begin{equation}
n_\mathrm{b}=u\, \left(n_\mathrm{ni}+n_\mathrm{pi}\right)+(1-u)\,n_\mathrm{no}+\mathcal A \nu_\mathrm{s}\, n_\mathrm{N}.
\label{nb}
\end{equation}
For fixed $X$, $R$   is fixed ($R\equiv [4\,\pi n_\mathrm{N}/(3u)]^{-1/3}$), but $\nu_\mathrm{s}$ depends on $\varepsilon_\mathrm{p}$ via $\mathcal A$.
However, the total number of neutrons adsorbed by the nucleus surface $N_s=\mathcal A \nu_\mathrm{s}$ is fixed (see Eq.\ \ref{nb}).

To present energy density in form (\ref{E}) we need to  separate energy for spherical nucleus for given $X$, which, as mentioned above, fixes  $N_\mathrm{s}$ and $R$.
For given $N_\mathrm{s}$ and $R$  surface density of adsorbed neutrons for spherical nucleus is $\nu_\mathrm{s}^X=N_\mathrm{s}/(4\pi R^2)$, thus respective surface energy is
\begin{equation}
E_\mathrm{s}^X(N_\mathrm{s},R)=4\pi R^2 \left[\mu_\mathrm{ns}(\nu_\mathrm{s}^X) \nu_\mathrm{s}^X +\sigma(\nu_\mathrm{s}^X)\right]
\end{equation} 
The surface energy of deformed nucleus can be written as
\begin{equation}
E_\mathrm{s}=E_\mathrm{s}^X(N_\mathrm{s},R)+\delta E_\mathrm{s},
\label{Es0+dEs}
\end{equation} 
where $\delta E_\mathrm{s}$ is a correction, associated with nucleus deformation, calculated at fixed number of adsorbed neutrons.
Introducing 
\begin{equation}
	\delta \nu_\mathrm{s}=\nu_\mathrm{s}-\nu_\mathrm{s}^X=-\frac{2}{5} \varepsilon_\mathrm{p}^2\nu_\mathrm{s}^X,
\end{equation}
and using Eq.\ (\ref{ThermCons_sigma}) we obtain
\begin{equation}
\delta E_\mathrm{s}=\frac{8}{5}\pi R^2\sigma \varepsilon_\mathrm{p}^2.
\label{deltaEs}
\end{equation} 

The next step is calculation of the Coulomb energy $E_\mathrm{C}(n_\mathrm{pi}, R,u,\varepsilon_\mathrm{p}, \varepsilon)$. It is essentially electrostatic problem.
Within considered CLDM, $E_\mathrm{C}$ is the Coulomb energy of a electrically neutral system (the cell), composed of uniformly positively  charged spheroid (nucleus) inserted into the center of larger spheroid with uniform negative  charge density (electron background).

As usually in electrostatics, we start from the Poisson's equation:
\begin{equation}
\Delta \varphi = -4 \pi \left[\rho_\mathrm{p} \Theta(R_\mathrm{sp}(\theta)-r)+\rho_\mathrm{e}\Theta \left(r_\mathrm{c}(\theta)-r\right)\right],
\label{Poisson}
\end{equation}
where $r$ and $\theta$ are the radial distance and the polar angle of the  spherical coordinate system ($r=0$ is the center of the cell),
$\Theta(x)$ is the Heaviside step function.
$\rho_\mathrm{p}=Ze/(4\pi R^3/3)=e n_\mathrm{pi}$  is the proton charge density inside nucleus, 
$\rho_\mathrm{e}=-u \rho_\mathrm{pi}$ is the electron charge density (the cell is electrically neutral).
The protons are located within a spheroid, with~boundary given by:
\begin{equation}
R_\mathrm{sp}(\theta) = \frac{R}{\sqrt{(1-\cos^2 \theta)(1+\varepsilon_\mathrm{p}) +\cos^2 \theta/(1+\varepsilon_\mathrm{p})^2 }}.
\end{equation}
The cell is also a spheroid, with~boundary given by:
\begin{equation}
r_\mathrm{c}(\theta) = \frac{a}{\sqrt{(1-\cos^2 \theta)(1+\varepsilon) +\cos^2 \theta/(1+\varepsilon)^2 }}.
\end{equation}

The solution to Equation~(\ref{Poisson}) is a sum of the proton potential $\varphi_\mathrm{p}$ and the electron potential $\varphi_\mathrm{e}$.
Namely, the electron potential inside the cell [$r<r_\mathrm{c}(\theta)$] is:
\begin{equation}
\varphi_\mathrm{e}=-2 \pi \rho_\mathrm{p}u a^2+\frac{2\pi}{3} \rho_\mathrm{p}u r^2 - \frac{4\pi}{5} \rho_\mathrm{p}u r^2 \varepsilon P_2(\cos \theta) + \frac{2\pi}{5} \rho_\mathrm{p}u a^2 \varepsilon^2 +\ldots,
\end{equation}
the proton potential inside the nucleus [$r<R_\mathrm{sp}(\theta)$] has similar form:
\begin{equation}
\varphi_\mathrm{p}=2 \pi \rho_\mathrm{p} R^2-\frac{2\pi}{3} \rho_\mathrm{p} r^2 + \frac{4\pi}{5} \rho_\mathrm{p} r^2 \varepsilon_\mathrm{p} P_2(\cos \theta) - \frac{2\pi}{5} \rho_\mathrm{p} R^2 \varepsilon_\mathrm{p}^2 +\ldots.
\end{equation}
Here $P_2(\cos \theta)=(3\cos^2 \theta-1)/2$ is the second Legendre polynomial.
The proton potential outside the nucleus is not required to calculate Coulomb energy (see below) and thus is not shown here.  The omitted  terms contributes to the Coulomb energy only at the third and higher order in infinitesimal parameters $\varepsilon$ and $\varepsilon_\mathrm{p}$, which is not considered here.

The total Coulomb energy can be presented as $E_\mathrm{C} = E_\mathrm{C}^\mathrm{pp}+E_\mathrm{C}^\mathrm{ep}+E_\mathrm{C}^\mathrm{ee}$,
where terms $E_\mathrm{C}^\mathrm{pp}$, $E_\mathrm{C}^\mathrm{ep}$, and~$E_\mathrm{C}^\mathrm{ee}$ are proton-proton, electron-proton, and~electron-electron 
contributions, respectively:
\begin{eqnarray}
E_\mathrm{C}^\mathrm{pp}&=& \frac{1}{2}\int_{r<R_{sp}(\theta)}\rho_\mathrm{p} \varphi_\mathrm{p}(\bm r)  d^3 \bm {r},
\\
E_\mathrm{C}^\mathrm{ep}&=&
\int_{r<R_{sp}(\theta)}\rho_\mathrm{p} \varphi_\mathrm{e}(\bm r)  d^3 \bm {r},
\\
E_\mathrm{C}^\mathrm{ee}&=&\frac{1}{2}\int_{r<r_{c}(\theta)}\rho_\mathrm{e}\varphi_\mathrm{e}(\bm r) d^3 \bm {r}.
\end{eqnarray} 
Here, we take into account that proton density is zero outside the spheroidal nucleus.

Analytic integration, accounting for the terms up~to the second order in $\varepsilon$ and $\varepsilon_\mathrm{p}$, leads to:
\begin{eqnarray}
E_ C^\mathrm{pp}&=& \frac{3}{5}\left(1 -\frac{1}{5} \varepsilon_\mathrm{p}^2\right)\frac{(Ze )^2}{R},
\label{Epp}\\
E_ C^\mathrm{ep}&=&
-\frac{3}{2} \frac{(Ze )^2}{a}\left(1-\frac{1}{5}\varepsilon^2\right)
 -\frac{9}{25}\frac{(Ze )^2}{R}u\varepsilon \varepsilon_\mathrm{p}
 \nonumber \\ 
 &&+\frac{3}{10} \frac{(Ze )^2}{R}u \left(1+\varepsilon_\mathrm{p}^2 \right),
 \label{Eep}\\
E_ C^\mathrm{ee}&=&
\frac{3}{5} \left(1-\frac{1}{5}\varepsilon^2\right)\, \frac{(Ze )^2}{a}
\label{Eee}
\end{eqnarray}
Coulomb energy for spherical nuclei in spherical cell corresponds to $\varepsilon=\varepsilon_\mathrm{p}=0$ and agrees with the well-known expressions (e.g., \citealt{hpy07}).


Presenting energy in form (\ref{E}) using (\ref{Es0+dEs}), (\ref{deltaEs}), and (\ref{Epp})--(\ref{Eee}) we identify explicit expressions for $f_1$, $f_2$, and $f_3$:
 \begin{eqnarray}	
f_1&=&\frac{9}{50}\frac{Z^2 e^2}{a},\\
f_2&=&\frac{8}{5}\pi R^2\sigma+\frac{3}{50}\frac{Z^2 e^2}{R}(5u-2),\\
f_3&=&-\frac{9}{25}\frac{Z^2 e^2}{R}u.
\end{eqnarray}

\label{lastpage}
\end{document}